\documentclass[aps,twocolumn,floats,superscriptaddress,showpacs]{revtex4}
\usepackage{amsmath,bm}%
\usepackage{graphicx}%

\begin{document}
\title{Directed and Elliptic Flow in $^{112}$Sn + $^{112}$Sn
Collisions below 100 MeV/nucleon
}

\author{H.Y. Zhang}
\affiliation{Shanghai Institute of Nuclear Research, Chinese
Academy of Sciences, Shanghai 201800, China}

\author{ Y.G. Ma}
\thanks{Email: ygma@sinr.ac.cn}
\affiliation{Shanghai Institute of Nuclear Research, Chinese
Academy of Sciences, Shanghai 201800, China}

\author{L.P. Yu}
\affiliation{Shanghai Institute of Nuclear Research, Chinese
Academy of Sciences, Shanghai 201800, China}

\author{W.Q. Shen}
\affiliation{Shanghai Institute of Nuclear Research, Chinese
Academy of Sciences, Shanghai 201800, China}

\affiliation{Ningbo University, Ningbo 315211, China}

\author{X.Z. Cai}
\affiliation{Shanghai Institute of Nuclear Research, Chinese
Academy of Sciences, Shanghai 201800, China}

\author{D.Q. Fang}
\affiliation{Shanghai Institute of Nuclear Research, Chinese
Academy of Sciences, Shanghai 201800, China}

\author{P.Y. Hu}
\affiliation{Shanghai Institute of Nuclear Research, Chinese
Academy of Sciences, Shanghai 201800, China}

\author{ C. Zhong}
\affiliation{Shanghai Institute of Nuclear Research, Chinese
Academy of Sciences, Shanghai 201800, China}

\author{ D.D. Han}
\affiliation{ East China Normal University, Shanghai 200062,
China}

\date{\today}

\begin{abstract}
     The directed and elliptic flow in collisions of
$^{112}Sn$ + $^{112}Sn$ at energies from 35 to 90 MeV/nucleon were
studied in an isospin-dependent quantum molecule dynamics model
(IQMD). With increasing incident energy, the directed flow rises
from the negative flow to the positive flow. Its magnitude depends
on the nuclear equation of state (EOS). However, the elliptic flow
shows decrease with increasing incident energy and its magnitude
is not very sensible to EOS. Systematic studies of the impact
parameter dependence and cluster mass dependence were also
performed. The study of directed flow at intermediate energies
thus provides a means to extract the information on the nuclear
equation of state.
\end{abstract}

\maketitle

Heavy ion collisions (HIC) provide a possibility to study the
properties of nuclear matter in conditions vastly different from
that in normal nuclei, such as high density and excitation as well
as large difference in the proton and neutron numbers. Such
knowledge is not only of interest in itself but also useful in
understanding astrophysics phenomena such as the evolution of the
early universe. One observable that has been extensively used for
extracting such information from heavy ion collisions is the
collective flow of various particles \cite{1}. The prediction of
collective flow in heavy ion collisions by the hydrodynamics model
\cite{2} has yielded a powerful tools for the investigation of
excited nuclear matter. The main goals are to determine the
nuclear equation of state (EOS) and the in-medium nucleon-nucleon
cross section \cite{3,4,5,6,7}. Recently, the isospin dependence
of collective flow has also become a very interesting subject of
theoretical and experimental studies \cite{8,9}. One knows that
nuclear collective flow is a kind of collective phenomenon found
in intermediate and high energy HIC, and the studies of the
dependence of collective flow on beam energy, mass number, and
impact parameter have revealed much interesting physics about the
properties and origin of collective flow. In past years, either
directed flow or the elliptic flow was studied separately in some
papers, but the combined researches are still very few. In this
work, an endeavor will be made along this direction.

The isospin dependence of collective flow has been studied
 by Li et al. \cite{10} and
Zheng et al. \cite{11} in term of an isospin-dependent
Boltzmann-Uehling-Uhlenbeck (IBUU) model in which the initial
proton and neutron densities were calculated from the nonlinear
relativistic mean-field (RMF) theory while the isospin dependence
enters the model by using the experimental N-N cross sections and
the isospin dependent nuclear mean field. However, the IBUU model
cannot describe the fragment flow physically since it is a
one-body transport model and does not contain many-body
correlation.

In the QMD model the nucleons are represented by Gaussian-shaped
density distribution. They are initialized in a sphere with a
radius R=1.12A$^{1/3}$, according to the liquid drop model. Each
nucleon is supposed to occupy a volume of $h^3$, so that the phase
space is uniformly filled. In order to explain some experimental
results, we have improved the original version of the QMD model
\cite{12} to include explicitly isospin degrees of freedom and get
an isospin-dependent QMD model (called IQMD model) which includes
isospin-dependent Coulomb potential, symmetry potential, N-N cross
sections, and Pauli blocking. In addition, in initialization of
projectile and target nuclei, we sample neutrons and protons in
phase space separately. Using the IQMD model, the directed and
elliptic flow in reaction of $^{112}Sn$+$^{112}Sn$ system in the
low-intermediate energy domian has been studied. Meanwhile, the
roles of the impact parameter and the nuclear equation of state
(EOS) have been also studied in this paper. One of the important
advantages of the IQMD model is that it can explicitly represent
the many body state of the system and thus contains correlation
effects to all orders. Therefore, the IQMD model provides
important information about both the collision dynamics and the
fragmentation process. Even though the BUU model can describe one
body observable very successfully, it fails to describe the
formation of clusters. In this paper, we construct clusters in
terms of the so-called coalescence model, in which particles with
relative momentum smaller than $P_0$ and relative distances
smaller than $R_0$ are considered to belong to one cluster. We
adopted the parameter $R_0$
 = 2.4 fm and $P_0$ = 200 MeV/c. In addition, in order to get rid of
nonphysical clusters, only the clusters with reasonable proton
number Z and neutron number N are selected. Taking the beam
direction along the z-axis and the reaction plane on the x-z
plane, the elliptic flow is then determined from the average
difference between the square of the x and y components of
particle transverse momentum \cite{13,14,15,16}, i.e.,
\begin{equation}
 V_2 = < \frac{P_x^2-P_y^2}{P_x^2+P_y^2}>
 \end{equation}

 It corresponds to the second Fourier coefficient in the
transverse momentum distribution \cite{17} and describes the
eccentricity of an ellipse-like distribution. In the intermediate
energy domain,  $V_2 > 0$  indicates of the in-plane enhancement
of particle emission, i.e., a rotation-like behavior, while $V_2 <
0$ characterizes the squeeze-out effect perpendicular to the
reaction plane, and $V_2 = 0$ means an isotropic distribution in
the transverse plane. Usually, $V_2$ is extracted from the
mid-rapidity region. The directed flow relates with the slope of
in-plane transverse momentum on the mid-rapidity in the C.M.
system, its  flow parameter $F$ at mid-rapidity is defined by
\begin{equation}
F = d<P_x/A>/dy
\end{equation}

\begin{center}
\begin{figure}
\includegraphics[scale=0.35]{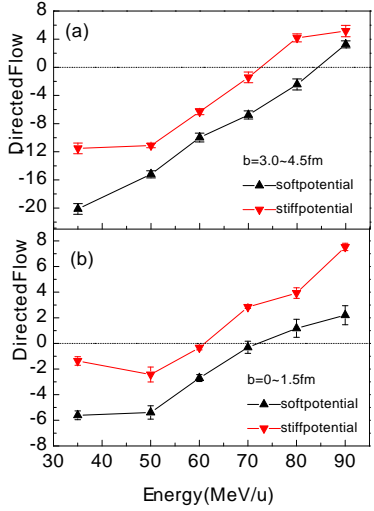}
\caption{\footnotesize Excitation function of the directed flow
with the soft or stiff EOS in different impact parameter zone.}
\end{figure}
\end{center}

The potential of the IQMD model used in the present study includes
an asymmetry term in the nuclear mean-field potential. The nuclear
mean-field potential is parameterized as
\begin{equation}
U(\rho,\tau_z) = a(\rho/\rho_0) + b(\rho/\rho_0)^\sigma + c \tau_z
(\rho_p-\rho_n)/\rho_0
\end{equation}

\begin{center}
\begin{figure}
\includegraphics[scale=0.35]{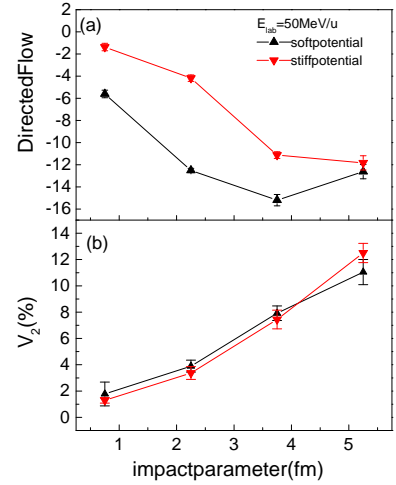}
\caption{\footnotesize  The impact parameter dependence of the
directed and elliptic flow with the soft or stiff EOS for 50
MeV/nucleon Sn + Sn.}
\end{figure}
\end{center}
In the above, $\rho_0$ is the normal nuclear density; $\rho$,
$\rho_n$, and $\rho_p$  are nucleon, neutron, and proton
densities, respectively; and $\tau_z$ equals 1 for proton and
$\tau_z$ = -1 for neutron and $c$ is chosen to be 32 MeV. The
parameter $a$ and $b$  was chosen to have a stiff equation of
state with a compressibility of K = 380 MeV and a soft equation of
state with a compressibility of K = 220 MeV as done in [12]. The
directed and elliptic flows in reaction $^{112}Sn$ + $^{112}Sn$
are calculated at different incident energies.  For each impact
parameter, a calculation of 200 events is performed. In the
present calculations, it is found that the directed and elliptic
flow have been saturated by the end of 120 fm/c, therefore the
results actually correspond to the statistical average value of
1000 "events" which come from the sum of five time intervals from
120 to 200 fm/c in each event.

Fig. 1 shows the directed flow for $^{112}Sn$ + $^{112}Sn$
collisions at different incident energies. Up-triangle symbols
correspond to soft potential and down-triangle symbols to stiff
potential. From Fig. 1, with increasing incident energy, the
directed flow increases. The balance energy, i.e., the energy of
disappearance of directed flow, for soft potential is larger than
that of stiff potential. From Fig. 1 (a), when the impact
parameter is 3.0$\sim$4.5fm, the balance energy with soft
potential is near 85MeV and that with stiff potential is near
70MeV. From Fig. 1 (b), when the impact parameter is 0$\sim$1.5fm,
the balance energy with soft potential is near 70 MeV and that
with stiff potential is near 60 MeV. Therefore, the balance energy
depends strongly on the nuclear equation of state as well as the
impact parameter. In this context, we can try to extract the
information on the nuclear equation of state from the excitation
function of the directed flow.

Fig. 2 shows the impact parameter dependence of the directed flow
(a) and the elliptic flow (b). The four impact parameter regions
0$\sim$1.5fm, 1.5$\sim$3.0fm, 3.0$\sim$4.5fm and 4.5$\sim$6.0fm
and incident energy 50MeV/nucleon are adopted in the calculation.
The directed flows are all negative while the elliptic flows are
all positive at different impact parameter. The directed and
elliptic flows depend strongly on impact parameter. With the
impact parameter increases, the directed flow increases in
negative direction and reaches the maximum in semi-peripheral
collision and decreases in negative direction in large impact
parameter. While, the elliptic flow always increases in positive
direction. These are consistent to the experimental observation
and other theoretical works \cite{18,19,20,21,22,23}. The directed
flow with soft potential and stiff potential have obvious
difference at different impact parameter, except for the
peripheral collision. However, the elliptic flow with soft
potential and stiff potential have almost the same values at all
impact parameters. It also means that the directed flow is a good
probe to pin down EOS than the elliptic flow in the studied system
and energy range .
\begin{center}
\begin{figure}
\includegraphics[scale=0.25]{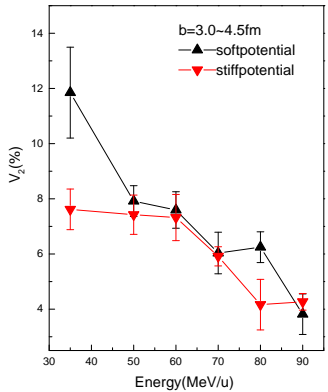}
\caption{\footnotesize Excitation function of the elliptic flow
with the soft or stiff EOS for Sn + Sn at b = 3 $\sim 4.5$ fm. }
\end{figure}
\end{center}

\begin{center}
\begin{figure}
\includegraphics[scale=0.35]{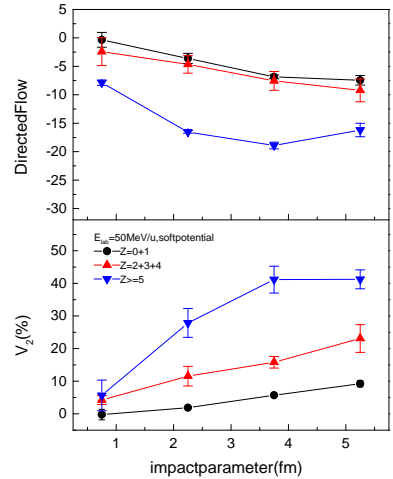}
\caption{\footnotesize Cluster mass dependence of the directed and
elliptic flow in different impact parameter.}
\end{figure}
\end{center}

When the impact parameter is smaller, the elliptic flow is
smaller, seen from Fig. 2(b), so a large impact parameter of
3.0$\sim$4.5fm is adopted in the Fig. 3 to present the excitation
function of elliptic flow. Fig. 3 shows the excitation function of
elliptic flow for $^{112}Sn$+$^{112}Sn$ collisions. All the values
of the elliptic flow are positive and decrease with the incident
energy, indicating the rotational behavior becomes weak with
increasing the incident energy, but the squeeze-out never reveal
below 90MeV/u even though the directed flow shows the change from
the negative to positive. When the incident energy is smaller, the
elliptic flow has large difference between the case of soft
potential and the case of stiff potential due to the fact that
nucleon-nucleon scattering effects at low energies are not strong
enough and the nuclear mean-field potential is dominated. But,
there will be only a little difference of the elliptic flow
between the soft potential and the stiff potential with the
incident energy increases. Overall speaking, the elliptic flow is
not very sensible to the nuclear equation of state in the studied
reaction system.

As said above, one of advantages of QMD is its fragment formation,
we show the directed and elliptic flow for some cluster
combinations in FIG. 4. The soft potential and the incident energy
50MeV/u are adopted in the FIG. 4. In order to accumulate enough
statistics, we define three bins of different fragments region
according to the charge, the solid circles are the flow of
neutrons and protons, the up-triangles are the flow of fragments
with Z=2$\sim$4 and the down-triangles are the flow of fragments
with Z$\geq$5. It is obvious to see that the absolute values of
the directed flow and the elliptic flow enhance with the charge
(or mass) increases, i.e. the heavier fragment flow is stronger
than the lighter one.

   In summary, the IQMD model has been used to study the directed and elliptic
flows in the collisions of $^{112}Sn$+ $^{112}Sn$ at different
impact parameters and different fragment charges from 35 to 90
MeV/nucleon. It is found that a transition in directed flow from
the negative to positive flow is revealed as the incident energy
increases. A strong dependence on the nuclear EOS is seen in the
directed flow at different incident energies and at different
impact parameters. The elliptic flow decreases with increasing the
incident energy and increases with increasing the impact
parameter. Meanwhile, the elliptic flow and the directed flow is
observed to be stronger as the fragment charge (mass) increases.
In comparison with the directed flow, the elliptic flow is not
sensitive to the EOS. By this study, we found that there exists
different sensitivity to EOS for different kind of flow. Hence, it
will be helpful to choose more sensitive flow probe to EOS before
the comparison with the data is made.

This work was supported in part by the NSFC for Distinguished
Young Scholar under Grant No. 19725521, the NSFC under Grant No.
19705012 and  the Major State Basic Research Development Program
of China under Contract No. G200077400.
{}

\begin{thebibliography}{}
\bibitem{1} Danielewicz P et al 1985 Phys. Lett. B {\bf 157} 146
\bibitem{2} Scheid W et al 1974 Phys. Rev. Lett. {\bf 32} 741
\bibitem{3} Xu H M 1992 Phys. Rev. Lett. {\bf 67} 2769
\bibitem{4} Ma Y G et al 1993 Phys. Rev. C {\bf 48} R1492
\bibitem{5} Ma Y G et al 1993 Z. Phys. A {\bf 344} 469
\bibitem{6} Reinsdorf W et al 1997 Annu. Rev. Nucl. Part. Sci. {\bf 47} 663 and reference therein
\bibitem{7} Tsang M B et al 1989 Phys. Rev. C {\bf 40} 1685
\bibitem{8} Pak R et al 1997 Phys. Rev. Lett. {\bf 78} 1022
\bibitem{9} Pak R et al 1997 Phys. Rev. Lett. {\bf 78} 1026
\bibitem{10} Li B A et al 1996 Phys. Rev. Lett. {\bf 76} 4492; Li B A 2000 Phys. Rev. Lett. {\bf 85} 4221
\bibitem{11} Zheng Y M et al 1999 Phys. Rev. Lett. {\bf 83} 2534
\bibitem{12} Aichelin J 1991 Phys. Rep. {\bf 202} 233¡¡
\bibitem{13} Appelshauser H et al 1998 Phys. Rev. Lett. {\bf 80} 4136
\bibitem{14} Danielenicz P et al 1998 Phys. Rev. Lett. {\bf 81} 2438
\bibitem{15} Shin Y et al 1998 Phys. Rev. Lett. {\bf  81} 1576
\bibitem{16} Sorge H 1999 Phys. Rev. Lett. {\bf 82} 2048
\bibitem{17} Voloshin S A 1997 Phys. Rev. C {\bf 55} R1630; Poskanzer A M et al 1998 ibid C {\bf 58} 1671
\bibitem{18} P\'eter J et al 1995 Nucl. Phys. A {\bf 519} 611
\bibitem{19} He Z Y et al 1996 Nucl. Phys. A {\bf 598} 248
\bibitem{20} Shen W Q et al 1993 Nucl. Phys. A {\bf 551} 333
\bibitem{21} Ma Y G et al 1993 Z. Phys. A {\bf 346} 285
\bibitem{22} Ma Y G et al 1995 Phys. Rev. C {\bf 51} 1029
\bibitem{23} Ma Y G et al 1995 Phys. Rev. C {\bf 51} 3256
\end{thebibliography}
\end{document}